# Magnetic and transport parameters of LSMO and YBCO/LSMO films deposited on sapphire substrates

Blagoy S. Blagoev[a*], Timerfayaz K. Nurgaliev[a], Vladimír Štrbik[b], Emil S. Mateev[a], Andrzej J. Zaleski[c]

[a]Institute of Electronics, Bulgarian Academy of Sciences, 72 Tzarigradsko chaussee blvd., 1784 Sofia, Bulgaria

[b]Institute of Electrical Engineering, Slovak Academy of Sciences, 9 Dubravska cesta str., 84104 Bratislava, Slovak Republic

[c]Institute of Low Temperature and Structure Research, Polish Academy of Sciences, 2 Okolna str., 50-950 Wroclaw, Poland

[*] Corresponding author: blago_sb@yahoo.com; tel. +359 2 9795 890; fax +359 2 9753201

**Keywords:** LSMO film; YBCO/LSMO bilayer; magnetic and electrical properties; intergrain conduction channels.

**Abstract** The $La_{0.7}Sr_{0.3}MnO_3$ (LSMO) layers and $YBa_2Cu_3O_{7-\delta}/La_{0.7}Sr_{0.3}MnO_3$ (YBCO/LSMO) bilayers were grown by magnetron sputtering on sapphire ($Al_2O_3$ or ALO) substrates. Temperature dependences of resistance of single LSMO films, grown on ALO substrates were typical for polycrystalline manganite materials and the resistance decreased with decrease of the temperature at medium temperatures and increased at lower and higher temperatures. Deposition of a top YBCO layer led to a drastic increase of the sample resistance. These bilayers did not demonstrate a decreasing of the resistance with decrease of temperature. Temperature dependence of the resistance of these samples was interpreted in the framework of a phenomenological model of two intergrain conduction channels. In framework of this model, parameters of the samples were determined and discussed.

## 1. Introduction

In last decades a transition-metal oxides provoke enormous increasing interest. These materials are attractive with their exotic properties such as high temperature superconductivity (HTS), colossal magnetoresistance (CMR), multiferroism and superparamagnetism [1-3]. The rich physics of these oxides is due to spin, charge and orbital degrees of freedom and their coupled dynamics [4-6]. A special attention is paid to perovskite manganese materials (or manganites) $L_{1-x}A_xMnO_3$ (L = La, Pr,…; A = Ca, Sr, Pb,…) which demonstrate a pronounced CMR effect (a decreasing of the electrical resistivity in external magnetic field) in ferromagnetic state near to the Curie temperature. In polycrystalline materials a low field magnetoresistance (LFMR) effect takes place, which is considered to be a result of spin polarized tunneling of charge carriers through the intergrain barriers [4, 5].

Thin films and heterostructures of transition-metal oxides manifesting HTS and CMR properties are specially promising for application in microelectronics and spintronics. The growing conditions of the film and the characteristics of the substrates affect strongly on the physical properties of thin films [4-6, 7-10]. Good quality epitaxial manganite films can be grown on conventional monocrystalline substrates, such as $SrTiO_3$, $LaAlO_3$ and MgO [4-6, 11-17]. Ferromagnetic thin manganite films can be prepared on unconventional substrates as Si and $Al_2O_3$ as well [18-21]. Heterostructures, consisting of manganite and HTS $YBa_2Cu_3O_{7-\delta}$ thin films, demonstrating excellent magnetic and superconductive properties, are successfully growing on conventional monocrystalline substrates (for example, on SrTiO$_3$, LaAlO$_3$ and MgO). Such layered structures were used in series of papers [22-25] for investigation of the fundamental problem of interaction of magnetism and superconductivity in the manganite/HTS interfaces and for device application tasks. At the present there is little a information in the literature [26-28] concerning characteristics of layered YBCO/manganite structures deposited on nontraditional substrates. Such kind of investigations are of great importance for understanding the physical processes in them and for possible device applications.

For this reason, in this paper we report the results of investigation of the resistive and magnetic properties of La$_{0.7}$Sr$_{0.3}$MnO$_3$ (LSMO) and YBa$_2$Cu$_3$O$_{7-\delta}$/LSMO (YBCO/LSMO) films deposited by magnetron sputtering on Al$_2$O$_3$ substrates.

## 2. Experimental details

Thin films La$_{0.7}$Sr$_{0.3}$MnO$_3$ (LSMO) and double YBa$_2$Cu$_3$O$_{7-\delta}$/La$_{0.7}$Sr$_{0.3}$MnO$_3$ (YBCO/LSMO) layers (or bilayers) were sputtered on monocrystalline Al$_2$O$_3$ substrates. The polycrystalline LSMO thin films of 40 nm were deposited by RF off-axis single magnetron sputtering on double-side polished 5×10×0.5 mm$^3$ r-cut sapphire (Al$_2$O$_3$ or ALO) substrates. The details of the deposition conditions can be found in our previous work [26]. An in-situ annealing at 500°C of substrate temperature and 600 Torr of oxygen pressure for 30 min took place for sample 1 and 3 while sample 2 was directly cooled to room temperature (RT) without annealing procedure (less oxygenated).

As a next step, YBCO films of 60 nm and 80 nm were sputtered by DC off-axis double magnetron system on the top of an obtained LSMO films: 1A (60nm/40nm), 2A (60nm/40nm - less oxygenated) and 3A (80nm/40nm). The standard deposition and annealing conditions were used for growing YBCO films [29]. Sample 3A with the thicker YBCO film was used for investigation of the characteristic magnetic properties of such kind of bilayers.

Magnetic properties of samples were investigated using PPMS® (Physical Property Measurement System) on Quantum Design. The DC magnetization measurements were performed by following ways: first, the sample was cooled without magnetic field from 300 K down to 4 K, afterwards a DC magnetic field of 100 Oe was applied and the measurements were made in increasing temperatures (ZFC-measurements); second, the magnetization was measured at applied magnetic field of 100 Oe at decreasing temperatures (FC-measurements); third, after FC-measurements the magnetic field was decreased to zero (in not overshoot mode) at 4 K and the measurements of the remanence magnetization were done at increasing temperatures (ReM-measurements). For ACM measurements the sample was first cooled down to 100 K without magnetic field and after that it was measured at AC field with amplitude 10 Oe and frequency 9967 Hz for increasing temperature. In all measurements the magnetic field was applied parallel to the substrate. The electrical resistance of the samples was measured using standard four probe method.

## 3. Results and discussions

### 3.1. Magnetic properties

The results of DC magnetization measurements of sample 3A (YBCO 80nm/LSMO 40nm) are shown in Fig. 1. The sample does not demonstrate any diamagnetic properties, caused by superconducting transition. This means that the YBCO layer of the YBCO/LSMO bilayer deposited on an ALO substrate remains in normal state at low temperatures. On the other hand, the sample reveals the magnetic properties (with the Curie temperature $T_{\text{Curie}}$=330 K) typical for manganite materials of polycrystalline nature. The FC magnetization increases monotonously with decrease of the temperature and reaches nearly a stationary level at $T$~200

K (Fig. 1). The ZFC magnetization, the value of which is small at low temperatures, increases with an increase of the temperature and manifests a maximum at $T\sim240$ K. At temperatures $T > T_{irr}\sim250$ K the ZFC branch of the magnetization curve coincides with that of the FC curve (where $T_{irr}$ is the temperature of irreversibility of the sample at magnetic field ~100 Oe). This type of irreversibility could indicate a presence of a nonuniform magnetic state in the LSMO film at low temperatures, which is similar to a cluster-spin-glass system with a characteristic spin-glass transition temperature $T_{irr}$. Such a magnetic disorder is hardly affected by the external magnetic field, so the irreversibility temperature and the behaviors of the FC magnetization will depend on the magnetic field as well (see for example FC curves in Fig. 11 from [18]).

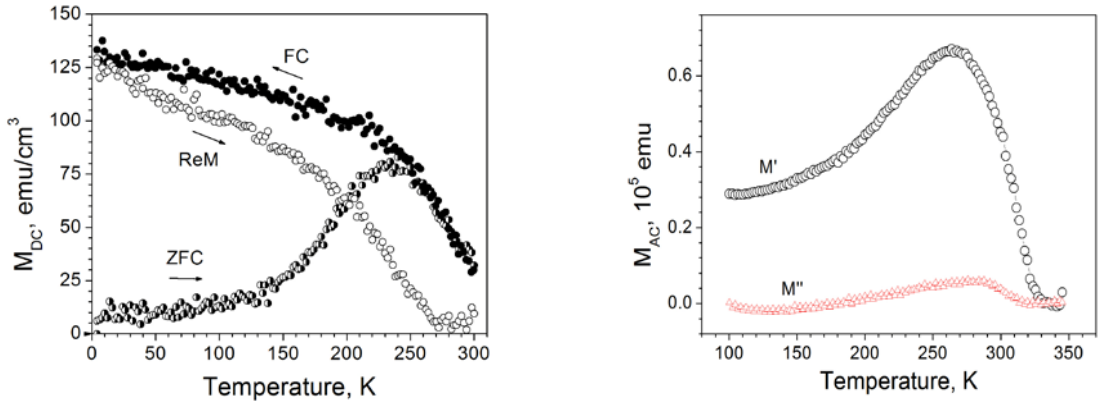

**Fig. 1** The zero field cooled (measured at 100 Oe), field cooled (measured at 100 Oe) and remanence magnetizations of YBCO/LSMO bilayer on sapphire substrate (sample 3A).

**Fig. 2** The real (in phase) and imaginary (out of phase) parts of AC differential complex magnetization of YBCO/LSMO bilayer on sapphire substrate (sample 3A).

With increase of the temperature the remanence magnetization ReM (induced at 4 K by the magnetic field of 100 Oe) decreases more rapidly than the FC magnetization and gets nearly zero at about 270 K which is greater than the irreversibility temperature and smaller than the Curie temperature of the sample. Between this temperature and the temperature of Curie, the total DC magnetization of the sample remains nearly zero despite the ferromagnetic nature of the sample.

Temperature dependences of the real $M'$ (in phase) and the imaginary $M''$ (out of phase) parts of AC magnetization of this sample 3A are shown in Fig. 2. The AC magnetization is small at low temperatures because of presence of "frozen" immobile magnetic moments in the sample. The "frozen" magnetic moments are melting with increase of the temperature and the real part of magnetization $M'$ increases and reaches maximum at about 270 K. At higher temperatures $M'$ decreases rapidly and disappears at $T= T_{Curie}= 330$ K. The imaginary $M''$ part of AC magnetization characterizes AC electromagnetic losses and it manifests a maximum at 280 K.

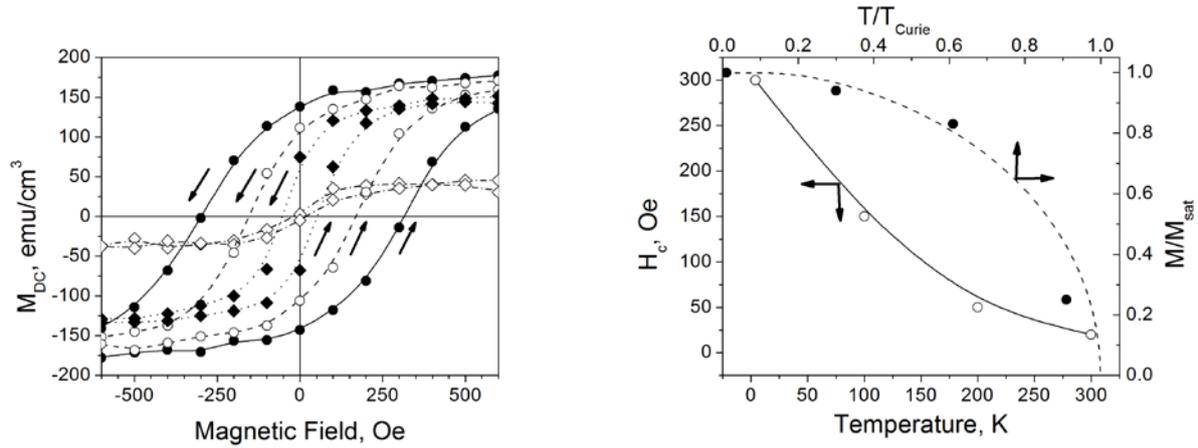

**Fig. 3** Fragments of the hystetesis curves obtained in field 2000 Oe for the sample 3A at 4 K (closed circles), 100 K (open circles), 200 K (closed diamonds) and 300 K (open diamonds).

**Fig. 4** Temperature dependences of the coercive field (open circles) and of the dimensionless magnetization measured at magnetic field of 1 kOe (closed circles – experiment, dashed line – results from calculation using formula (2) and parameters where: $p = 3$; $M_{sat} = 180$ emu/cm$^3$; $T_{Curie}= 330$ K for sample 3A.

Hysteresis loops of the sample 3A measured at 4, 100, 200 and 300 K are shown in Fig. 3. They do not manifest any indications of presence of superconducting properties of this bilayer too. The width of hysteresis loop (the coercive field $H_c$) decreases nearly exponentially with increasing of the temperature (Figs 3 and 4). The temperature dependence of the magnetization can be described approximately using the Brillouin function (formula (2)). The saturation magnetization of the sample is several times smaller than that of monocrystalline bulk LSMO samples.

### 3.2. Electrical resistance

Temperature dependences of resistance of single LSMO films, grown on ALO substrates (samples 1 and 2) are shown in Figs 5 and 6. The experimental curves are typical for polycrystalline manganite materials and films [30-34]. It is known that LSMO films deposited by different methods are grown polycrystalline on sapphire substrate [35-37].

The electrical resistance initially rises, attaining a maximum at the metal-insulator transition temperature ($T_{MI} \sim 170$ K and 210 K for samples 1 and 2 respectively). Below $T_{MI}$, down to ~55 K, the behavior is changed to a metallic-like which is characteristic for a ferromagnetic-metallic phase. Below 55 K the resistivity increases again suggesting that grain boundary scattering plays an important role in the transport properties of the films.

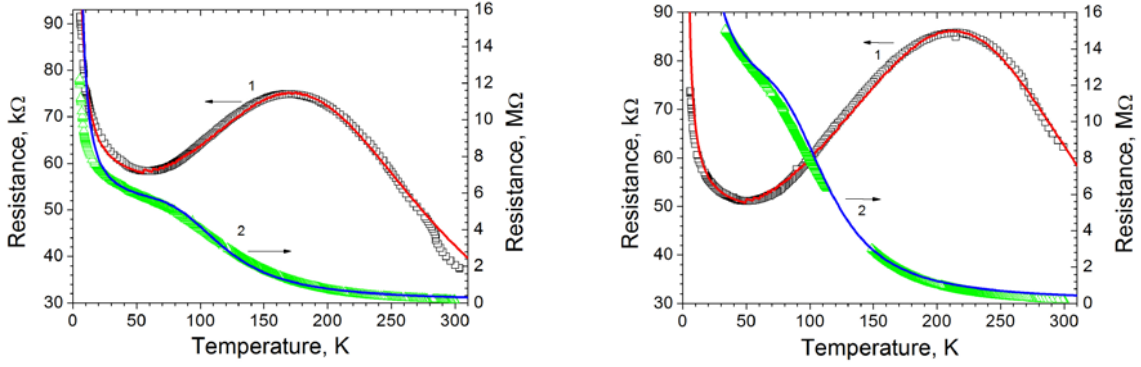

**Fig. 5** Experimental (open squares and triangles) and modeled (solid lines) *R vs T* dependences of LSMO film (sample 1, dependence 1) and YBCO/LSMO bilayer (sample 1A, dependence 2) deposited on ALO substrates. Parameters, used in modeling, are given in Table 1.

**Fig. 6** Experimental (open squares and triangles) and modeled (solid lines) *R vs T* dependences of LSMO film 2 (dependence 1) and bilayer 2A (dependence 2) deposited on ALO substrates. Parameters, used in modeling, are given in Table 1.

The temperature dependences resistance of YBCO/LSMO bilayers deposited on ALO substrates (samples 1A and 2A) are shown in Figs 5 and 6 as well. These dependences do not manifest features, characteristic for a superconducting transition, and the samples show a high value of the resistance. Structural and dimensional mismatches [38] between LSMO and ALO stimulate a growth of polycrystalline LSMO films with a large number of grain boundaries and small grain sizes (see also [31, 32]). The surfaces of the grains and grain boundaries may be contaminated because of difference of the ions environment at the surface and the inside of the grains. For this reason, the growth of superconducting YBCO crystalline phase is hindered on such a surface. High resistive non-superconducting YBCO top layer leads to modification of grain boundaries in LSMO film and to a modification of the temperature dependence of the resistance as it is observed in our samples (Figs 5 and 6).

Temperature dependence of the resistance of the samples were interpreted using a phenomenological model of intergrain conduction channels. This model was developed in [30-33, 39] for description of the electrical resistivity behavior in granular magnetic materials. In the framework of this model one kind of the channel reflects the transport properties of the system achieved through spin-polarized tunneling between neighboring ferromagnetic grains with good contacts. The other kind of conduction channel is achieved through thermal activation of the insulator phase derived from the grain boundary, contaminations, crystal disorder and poor connectivity between grains. The measured resistance *R* is determined by the total conductance of above two channels:

$$R^{-1} = R_0^{-1}\left\{a_1 e^{-W_2/(2k_B T)} + a_2 e^{-[E_c + W_1(1-m^2)]/(2k_B T)}\right\}, \tag{1}$$

where $R_0$ is a constant related to the sample resistance, $a_2$ and $a_1$ are dimensionless effective section constants for the insulator phase channels (modeled by a semiconductorlike resistance) and the channel, responsible for spin-polarized tunneling, respectively, $k_B$ is Boltzmann's constant, $T$ is the absolute temperature, $W_2$ is the insulator phase activation energy, $E_C$ – charging energy, $W_1(1-m^2)$ is the magnetic correlation function of two neighboring grains of the same volume and shape, $W_1$ – temperature and magnetic field independent constant and $m$ is the dimensionless magnetization normalized to the saturation value for each grain.

For modeling of the temperature dependence of the grain magnetizations we successfully used the Brillouin function $B_p(\alpha)$ [40]

$$m = \frac{I}{I_0} = B_p(\alpha), \qquad \alpha = \frac{I_0 H}{N k_B T} + \frac{3p}{p+1}\left(\frac{T_{Curie}}{T}\right)\left(\frac{I}{I_0}\right), \qquad (2)$$

where $H$ is the external magnetic field, $T_{Curie}$ is Curie temperature, $I_0$ is saturation magnetization at 0 K, $I$ is magnetization and $N$ is the number of magnetic atoms per unit volume. Figures 5 and 6 (curves 1) show the modeled $R$ vs $T$ curves together with the experimental dependences for the single LSMO layers deposited on ALO substrates (samples 1 and 2). As fitting parameters were used $R_0$, $a_1$, $a_2$, $W_1$, $W_2$, $E_C$, $T_{Curie}$, $p$ and their magnitudes are given in Tab.1. It is seen that the calculated curves satisfactory match the experimental data. The extracted Curie temperature $T_{Curie}$ of 340 K for the samples is close to the one (330 K) obtained from AC magnetization measurements. In the case of a less oxygenated sample 2, the effective section constant $a_1$ for the spin - polarized tunneling channels is smaller than that obtained in the sample 1. Such a less oxygenated sample is characterized with higher values of the activation energies of the charge carriers in the both channels of the conductivity as well. It can be noted also, that parameters $a_1/a_2$, $W_1/2k_B$, $W_2/2k_B$, $E_C/2k_B$, obtained for our LSMO films grown on ALO substrate, are not substantially different from those obtained for thin film LSMO composites grown on Si (111) substrates [30].

Table 1. Fitting parameters.

| Sample | $R_0$, (Ω) | $a_1$ | $a_2$ | $W_1/2k_B$ (K) | $W_2/2k_B$ (K) | $E_C/2k_B$ (K) | $p$ | $T_{Curie}$ (K) |
|---|---|---|---|---|---|---|---|---|
| 1 | 7.295 10$^3$ | 0.0405 | 0.9595 | 390 | 980 | 3 | 1.9 | 340 |
| 1A | 5.05 10$^6$ | 0.0099 | 0.9901 | 100 | 580 | 9 | 1.9 | 340 |
| 2 | 1.8527 10$^4$ | 0.0285 | 0.9715 | 442 | 1310 | 2.5 | 2.4 | 340 |
| 2A | 1.61 10$^7$ | 0.0062 | 0.9938 | 100 | 620 | 15 | 2.4 | 340 |

Deposition of top YBCO layer leads to a drastic increase of resistance of the samples (see Fig. 5, 6, dependences 2) due to high resistivity of the non-superconducting top YBCO layer and to a modification of the grain boundaries. As LSMO and YBCO films are electrically "connected" in parallel in the sample, the temperature dependence of the resistance was simulated using the above model as well. Modeled temperature dependences of the resistance for these samples 1A and 2A (the model parameters are given in Tab.1) are shown in Fig. 5 and 6. Very small values of the ratio of effective section constants $a_1/a_2$ for these samples (Tab.1) reflect a strong suppression of the channels responsible for a spin-polarized tunneling in LSMO layers of the sample. For this reason the samples 1A and 2A do not demonstrate a decreasing of the resistance with decrease of temperature which is characteristic for the samples with a spin-polarized tunneling. Both channels of conductivities of these bilayers are characterized by smaller values of activation energies $W_1/2k_B$, $W_2/2k_B$ for the charge carriers (in comparison with those of a single LSMO layer), which probably means, that the grain boundaries with higher activation energies (which could be considered as grain boundaries with lower qualities) completely lose their conductivity during deposition of the top YBCO layer. Charging energy $E_C$ is higher for these highly resistive films and it leads to a step like behavior of the $R$ vs $T$ dependences at low temperatures (see Fig. 5, 6, curves 2).

## 4. Conclusions

The La$_{0.7}$Sr$_{0.3}$MnO$_3$ (LSMO) layers and YBa$_2$Cu$_3$O$_{7-\delta}$/La$_{0.7}$Sr$_{0.3}$MnO$_3$ (YBCO/LSMO) bilayers were grown on sapphire (Al$_2$O$_3$) substrates. The layers were not monocrystalline.

Magnetic properties of the YBCO/LSMO bilayer indicated a presence of a nonuniform magnetic state in the LSMO film at low temperatures $T < T_{irr} \sim 250$ K.

Temperature dependences of resistance of single LSMO films, grown on ALO substrates were typical for polycrystalline manganite materials and the resistance decreased with decrease of the temperature at medium temperatures and increased at lower and higher temperatures. Deposition of a top YBCO layer on these samples led to a drastic increase of their resistance. Temperature dependence of the resistance of these LSMO films and YBCO/LSMO bilayers was interpreted in the framework of a phenomenological model of two intergrain conduction channels. Very small values of the ratio $a_1/a_2$ of effective section constants $a_1$, $a_2$ of these channels in YBCO/LSMO samples, determined in the framework of the above model, reflected a strong suppression of the conductivity of channels responsible for a spin-polarized tunneling in LSMO layers of the double-layer sample.

**Acknowledgements**
This work was supported by the European Social Fund under the project BG051PO001.3.3.04/54/2009 and by the Slovak Grant Agency VEGA under projects No. 2/0173/13 and 2/0164/11.